\documentclass[12pt]{iopart}
\usepackage{latexsym}
\usepackage{amssymb}
\usepackage{graphicx}
\usepackage{dcolumn}
\usepackage{bm}
\usepackage{xcolor}
\usepackage{float}

\newcommand{\mZ}{{\mathbb Z}}
\newcommand{\mN}{{\mathbb{N}}}

\newcommand{\half}{{{\textstyle\frac{1}{2}}}}

\newcommand{\be}{\begin{equation}}
\newcommand{\ee}{\end{equation} }
\newcommand{\beqa}{\begin{eqnarray} }
\newcommand{\eeqa}{\end{eqnarray} }
\newcommand{\ba}{\begin{array}}
\newcommand{\ea}{\end{array}}

\newcommand\rd{{\rm d}}

\newcommand\cO{{\cal O}}

\newcommand\cT{{\cal T}}
\newcommand\cV{{\cal V}}
\newcommand\cZ{{\cal Z}}

\newcommand\BEC{{\rm\scriptscriptstyle{BEC}}}

\newcommand\kB{k_{{\scriptscriptstyle{\rm B}}}}
\newcommand\dis{\displaystyle}

\newcommand\shp{{\ast\ast}}

\newcommand\Tp{{\cT}_{{\scriptscriptstyle{P}}}}
\newcommand\Vp{{\cV}_{{\scriptscriptstyle{P}}}}
\newcommand\Trho{{\cT}_{{\rho}}}

\newcommand\vn{\vec{n}}

\begin{document}

\title[]{Two-dimensional  Bose-Einstein condensate  under  pressure}

\author{Wonyoung Cho$^{\dagger}$, Sang-Woo Kim$^{\sharp}$ and Jeong-Hyuck Park$^{\dagger}$}

\address{$^{\dagger}$Department of Physics,  Sogang University, Seoul 121-742, Korea}
\address{$^{\sharp}$Korea Institute for Advanced Study, Seoul 130-722, Korea}
\address{$^{\sharp}$Institute of Theoretical Physics Chinese Academy of Sciences,   Beijing 100190, China}

\ead{park@sogang.ac.kr}

\begin{abstract}
Evading the Mermin-Wagner-Hohenberg no-go theorem and revisiting with rigor   the  ideal Bose  gas  confined  in a square box, we  explore   a discrete phase transition in two spatial dimensions.   Through both analytic and numerical methods  we verify  that  thermodynamic instability \textit{emerges}  if the number of particles is sufficiently  yet finitely large:  specifically $N\geq 35131$.  The instability implies that  the isobar of the  gas zigzags on the temperature-volume plane, featuring  supercooling  and  superheating phenomena.  The Bose-Einstein condensation  then can  persist   from absolute zero to the superheating  temperature.  Without   necessarily taking the large $N$ limit,  under constant pressure condition, the condensation takes place discretely   both  in    the momentum and  in  the  position spaces.   Our result is applicable to  a harmonic trap. We assert  that  experimentally observed Bose-Einstein  condensations of harmonically trapped  atomic gases are a  first-order phase transition which involves  a discrete change of the density    at the center of the trap. 
  
\end{abstract}

\vspace{2pc}
\maketitle

\section{Introduction}
The existence of Bose-Einstein condensate (BEC)  in two spatial dimensions ($2D$) is a  subtle  issue,  attracting  a wide range of interests   from  both theoretical and experimental perspectives.  The Mermin-Wagner-Hohenberg (MWH)  theorem~\cite{MerPRL,HohPR} prohibits  free bosons from condensing   on a homogeneous  infinite  plane  \textit{via}  long-range thermal fluctuations.   The Berezinskii-Kosterlitz-Thouless (BKT) mechanism  then  provides  an alternative  explanation of  a quasi-transition to superfluid without condensation~\cite{BerSPJ,KosJPC}.  On the other hand,       $2D$  BEC has been    realized  in  experiments    for harmonically trapped atomic gases, \textit{c.f.~}\cite{HadzibabicDalibard} and  references therein,  in support of some analytic  estimations~\cite{BagnatoKleppner,Mullin97,Colloquium}. 

One  related  deep question is whether a finite system can feature a mathematical singularity or not.   While Anderson once stated  ``\textit{More is different}''~\cite{Anderson},  the  question we are  addressing  is      ``\textit{How many is different?}''~\cite{Park:2013gpa}.  Constrained by  the analyticity   of the partition function~\cite{YangLee} one may argue  that  ``\textit{More is the same: infinitely more is different}"~\cite{Kadanoff}. This has motivated, especially for $3D$,  the widely adopted  rather axiomatic approach to discrete phase transitions  where the first-order  phase transition is `defined' only in the thermodynamic limit.  In two and one dimensions,  the MWH theorem then seems to imply that \textit{Infinity is not enough: interaction is required}.  However, it is also true that  infinity is hardly realistic and exists only in theory.

Historically,  London already noted   in 1954~\cite{London}   that in order to decide the question of  the order of the transition   one should not  prescribe the volume, $V$; one has rather to use pressure, $P$, and temperature, $T$ as independent variables.  The specific heats per particle at constant volume, $c_{V}$, and constant pressure, $c_{P}$,  can be  quite  different  for a finite system from the theoretical consideration of statistical physics. Being  an analytic function, $c_{V}$ should be  finite for a finite system. On the other hand,  $c_{P}$ is generically given by a fraction of two analytic functions and when the  denominator vanishes, namely on the spinodal curve~\cite{spinodalnuclear,Kardar,Yukalov},  it  becomes singular, for further discussions  see \textit{e.g.~}\cite{Imry,ChabaPathria} as well as  Eq.(6) of \cite{7616PRA}.

   Essentially,  one can easily  boil  water if one keeps not the volume (or density)  but the pressure constant~\cite{7616PRA,Park:2013gpa}.  Indeed,   involving two of the authors,  it has been  shown  recently  that  the $c_{P}$ of $3D$ ideal Bose gases  can diverge even for a finite number of particles~\cite{7616PRA,7616relativistic,Jeon:2011nk}. This is only  possible   because  thermodynamic instability emerges  making   the isobar zigzag on the $(T,V)$-plane  when  the number of particles is greater than or equal to a definite  value. Such critical number depends on the shape of the box but not the size,  thanks to the existing scale symmetry of the ideal gas: for example the critical number is precisely  $7616$ for the $3D$ cubic box~\cite{7616PRA}. This might appear  mathematically unnatural and random, but can be a   \textit{physical} answer to the question,  \textit{How many is different?}.    Because the isobar zigzags on the $(T,V)$-plane, when  the temperature increases under constant pressure, the volume and hence the density  must make a `discrete' jump. Since all the physical quantities  are functions of temperature and density, the discrete jump then implies or realizes  a first-order phase transition.

In this work,  we turn  our attention to a $2D$ ideal Bose gas which is confined in a square box.  In contrast to the homogeneous infinite plane for which the MWH  no-go theorem surely holds, the presence of the box breaks the translational symmetry and converts the energy  spectrum from   continuous \textit{uncountable} infinite    to discrete  \textit{countable}  infinite. This leads to   fundamental differences between  the two systems, effectively   classical \textit{v.s.} quantum,   which should  persist  even in  the  large $V$  limit of  the box.  As the harmonic potential allows  $2D$  BEC to occur,   it is   physically  natural to expect  that a ``small''  box should  do so as well.

In fact, it has been known from 1970s that  $2D$ BEC may occur  under the constant pressure condition~\cite{Imry,ChabaPathria}.  The critical temperature was first obtained by Imry \textit{et al.} in \cite{Imry} and its leading order finite-size correction   was      analyzed by Chaba and Patria in  \cite{ChabaPathria}.  In particular, by insisting   the thermodynamic stability,   the latter two authors found  a pair of  nearby critical temperatures. Yet, as they remarked  in their  section III, the physical interpretation was somewhat ``awkward" and ``more refined treatment of the problem"  was   desirable.

 It is the purpose of  the present paper  to provide such a  desired    complementary  analysis,   helped by   modern computing power.     Treating the exact mathematical expressions  numerically --which all  descend  from a single  partition function-- we  show  that thermodynamic instability,   satisfying the spinodal condition, \textit{emerges}  if the number of  particles is sufficiently yet finitely large. Specifically for  the $2D$ ideal Bose gas confined in a square box, we demonstrate   that the isobar     zigzags  on the temperature-volume plane if  $N\geq N_{c}=35131$. Consequently,  the $2D$ BEC becomes a first-order  phase transition under constant pressure,  similar to    the   $3D$ ideal Bose gases~\cite{7616PRA,7616relativistic,Jeon:2011nk}.   We  identify the two critical points found in  \cite{ChabaPathria}  as the supercooling and the superheating spinodal points which correspond to the two turning points of the zigzagging isobar.   Between the supercooling and the superheating temperatures the volume is triple-valued including one unstable configuration.  We also improve  the analytic approximation  of \cite{ChabaPathria} by newly obtaining    `double logarithmic' sub-leading corrections.

 Our main   results are  spelled  in   Eqs.(\ref{MAINsupercooling},\ref{MAINsuperheating}) and depicted  in     Figures~\ref{FIGspinodal},\,\ref{FIGisobar},\,\ref{FIGd2bec}, which    are  methodologically    twofold: \textit{i)}  analytical for large $N$ and \textit{ii)}  numerical for arbitrary $N$. They  are in  excellent agreement,   for which  the double logarithmic terms are necessary.  While much of the present $2D$  results are parallel to the $3D$ cases~\cite{7616PRA,7616relativistic,Jeon:2011nk}, there are also  some differences.  Lastly,  we comment  on the implication  of our results to the experimentally observed BEC of harmonically trapped  atomic gases.\\
  
  \newpage

\section{Setup}
We consider  the textbook  $2D$   system of an ideal Bose gas which is confined in a square box,  with the area (or  $2D$  volume), $V$, and the number of particles, $N$. Experimental realization of such a system  has   just begun    this year~\cite{14064073}.  In a parallel manner to Ref.\cite{Jeon:2011nk} (\textit{c.f.~}\cite{3Dharmonic}), here  we focus on the grand canonical ensemble  with the fixed average number of particles. From the  non-relativistic dispersion relation,  $E={\vec{p}{}^{\,2}}/{(2m)}$,   the  grand canonical partition function    is actually    a two-variable function depending on  the fugacity, $z$,  and  the  combination  of the  temperature and the area, $TV$. This implies a scaling symmetry:   different sizes  of the volume can be traded with   different scales  of the temperature. In particular, the small volume or the `confining'  potential effect should persist for  large volume. 

Specifically  we set,  as for the two dimensionless    fundamental variables in our analysis,
\be
\ba{ll}
\varepsilon:= \frac{\pi^{2}\hbar^2}{2m\kB TV}\,,~~&~~
\sigma:=-\ln z\,.
\ea
\label{fundamental}
\ee
In terms of these, the grand canonical partition function reads
\be
\ln\cZ(\varepsilon,\sigma)=-{\sum}_{\vn\in\mN^2}\,\ln\!\left(1-e^{-\varepsilon\vn^{2}-\sigma}\right)\,.
\label{logZ}
\ee
With the Dirichlet boundary condition which we deliberately  impose, $\vn=(n_{1},n_{2})\in\mN^{2}$ is a positive integer-valued  $2D$  lattice vector, such that the lowest value of $\vn^{2}$ is  two  and $\sigma$ is bounded from below as $\sigma> -2\varepsilon$.

The average  number of particles is then
\be
N(\varepsilon,\sigma)=-\partial_{\sigma}\ln\cZ(\varepsilon,\sigma)\,,
\label{Nexpression2}
\ee
and  the standard expression,  $P=\kB T \partial_{V}\ln\cZ(T,V,z)$,  of the pressure  is equivalent to\footnote{In $2D$, the pressure  assumes the dimension,  ${\rm \left[P\right]= \left[\frac{force}{length}\right]=\left[\frac{mass}{~time^{2}}\right] }$.}
\be
\Tp(\varepsilon,\sigma):=\!\left(\frac{2m}{\pi^{2}\hbar^{2}P}\right)^{\frac{1}{2}}\!\kB T =\left[-\varepsilon^{2}\partial_{\varepsilon}\ln\cZ(\varepsilon,\sigma)\right]^{-\frac{1}{2}}.
\label{defTp}
\ee
Being a combination of\,  $T$ and $P$,  this  dimensionless quantity, $\Tp$,   can denote  the physical   temperature   on an arbitrarily   given isobar. Similarly we may define  a dimensionless ``volume",
\be
\Vp(\varepsilon,\sigma):=\left(\frac{2m}{\pi^{2}\hbar^{2}}P\right)^{\frac{1}{2}}V
=\left[-\partial_{\varepsilon}\ln\cZ(\varepsilon,\sigma)\right]^{\frac{1}{2}}\,,
\label{defVp}
\ee
and another dimensionless ``temperature",
\be
{\Trho(\varepsilon,\sigma):=\frac{2m\kB TV}{\pi^{2}\hbar^{2}N}
=\left[-\varepsilon\partial_{\sigma}\ln\cZ(\varepsilon,\sigma)\right]^{-1}\,.}
\label{defTrho}
\ee
Further, the number of  particles on the ground state is 
\be
{
N_{0}(\varepsilon,\sigma)=\partial_{\sigma}\ln\!\left(1-e^{-2\varepsilon-\sigma}\right)=
\left(e^{2\varepsilon+\sigma}-1\right)^{-1}\,.}
\label{N0}
\ee
As we   already denoted, $N$, $N_{0}$,  $\Tp$, $\Vp$ and   $\Trho$   are all  functions of the two variables, $\varepsilon$, $\sigma$ only. They  satisfy  identities,
\be
\ba{ll}
\Tp(\varepsilon,\sigma)\Vp(\varepsilon,\sigma)=\varepsilon^{-1}\,,\quad&\quad\Trho(\varepsilon,\sigma)N(\varepsilon,\sigma)=\varepsilon^{-1}\,.
\ea
\ee

Generically, the superheating (BEC)  and  the supercooling points correspond  to the  two turning points of an   isobar which zigzags on the $(T,V)$-plane,   satisfying the \textit{spinodal curve} condition,  ${\rd N}=0$, ${\rd P}=0$, ${\rd T}=0$~\cite{Kardar,spinodalnuclear,Yukalov,7616relativistic,Jeon:2011nk}. In our case,  the spinodal curve  is to be   positioned  on the $(\varepsilon,\sigma)$-plane  (\textit{c.f.~}Figure~\ref{FIGspinodalraw}) to satisfy 
\be
\ba{ll}
{\rd N(\varepsilon,\sigma)=0}\,,\quad&\quad
{\rd \Tp(\varepsilon,\sigma)=0}\,,
\ea
\ee 
and hence  the following linear equation must admit   a nontrivial solution,
\be
\left(\ba{c}0\\0\ea\right)=\left(\ba{cc}
\partial_{\varepsilon}\partial_{\sigma}\ln\cZ&\partial_{\sigma}^{2}\ln\cZ\\
\left(2\varepsilon^{-1}\partial_{\varepsilon}+\partial_{\varepsilon}^{2}\right)\ln\cZ~&\partial_{\varepsilon}\partial_{\sigma}\ln\cZ\ea
\right)\left(\ba{c}\rd \varepsilon\\\rd\sigma\ea\right)\,.
\label{lineareq}
\ee
Consequently  the ${2\times 2}$ matrix must be singular,
\be
\Phi:=\det \left(\ba{cc}
\partial_{\varepsilon}\partial_{\sigma}\ln\cZ&\partial_{\sigma}^{2}\ln\cZ\\
\left(2\varepsilon^{-1}\partial_{\varepsilon}+\partial_{\varepsilon}^{2}\right)\ln\cZ~&\partial_{\varepsilon}\partial_{\sigma}\ln\cZ\ea
\right)\equiv 0\,.
\label{Phispinodal}
\ee
This algebraic equation determines    the spinodal curve  on the $(\varepsilon,\sigma)$-plane. For consistency, we note   that the determinant, $\Phi$,  is  proportional to   $\left.\frac{\rd \Tp}{\rd \Vp}\right|_{N}$,  as
\be
{\left.\frac{\rd \ln\Tp}{\rd \ln\Vp}\right|_{N}=\frac{\Phi}{\,\left(\partial_{\varepsilon}^{2}\ln\cZ\right)
\left(\partial_{\sigma}^{2}\ln\cZ\right)-
\left(\partial_{\varepsilon}\partial_{\sigma}\ln\cZ\right)^{2}
}\,,}
\ee
where the denominator on the right hand side can be shown to be  positive definite~\cite{Jeon:2011nk}. Hence, the vanishing of the determinant  is,  as expected, equivalent to the vanishing of  $\left.\frac{\rd \Tp}{\rd \Vp}\right|_{N}$, implying a   zigzagging isobar (see Figure~\ref{FIGisobar}).

Our aims are  first  to solve  (\ref{Phispinodal}), second to  express the solutions in terms of the more physical variables, $N$, $\Tp$,  $\Vp$, $\Trho$   using (\ref{Nexpression2}), (\ref{defTp}), (\ref{defVp}), (\ref{defTrho}), and finally to confirm  BEC, \textit{i.e.~}$N_{0}/N\rightarrow 1$, at the superheating point.  Searching  for  spinodal curves approaching  the large $N$  limit,  we shall focus on  the region,  $\varepsilon\rightarrow 0^{+}$ and  $\sigma+2\varepsilon\rightarrow 0^{+}$.

\ref{AppendixA} contains our  technical yet self-contained derivation of the solutions which we spell below.\\

\newpage

\section{Main Results}
\begin{figure}[H]
\centering\includegraphics[width=100mm]{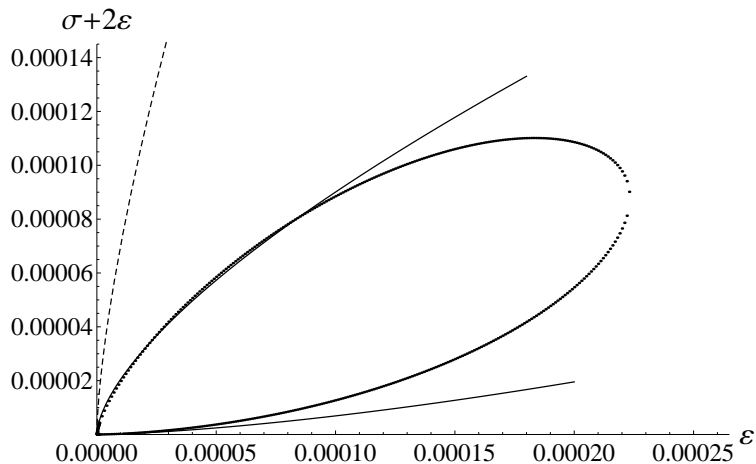}
\caption{The supercooling    and the superheating    spinodal curves  on the $(\sigma+2\varepsilon,\varepsilon)$-plane.
The dotted closed curve   is  from the numerical computations based on the exact formula (\ref{Phispinodal}). The upper and lower solid lines correspond to our  analytic approximate solutions,   (\ref{supercoolse}) and (\ref{superheatse}) respectively. The dashed line denotes a less-accurate supercooling  approximation~\cite{ChabaPathria} without the double logarithmic term in  (\ref{supercoolse}). The interior of the dotted closed curve is  thermodynamically unstable.   }
\label{FIGspinodalraw}
\end{figure}

Our solutions of the spinodal curves  solving the constraint (\ref{Phispinodal}) on the $(\varepsilon,\sigma)$-plane  are  are  depicted in Figure~\ref{FIGspinodalraw}. Especially near to the region,  $\varepsilon\rightarrow 0^{+}$,  $\sigma+2\varepsilon\rightarrow 0^{+}$ (as for large $N$), we obtain the following analytic approximate  solutions.

\begin{list}{}
\item $\ast$ Supercooling spinodal curve of $h=\frac{1}{2}$,
\be
\ba{cll}
\sigma&\simeq&-2\varepsilon+{\frac{2\zeta(\frac{3}{2})}{\sqrt{\pi}}\left(\ln\varepsilon-4\ln|\ln\varepsilon|
\right)^{-2}\varepsilon^{\frac{1}{2}}\,,}\\
\varepsilon&\simeq&{\frac{\pi}{8}}N^{-1}\left(\ln N +3\ln\ln N\right)\,.
\ea
\label{supercoolse}
\ee
\item $\shp$  Superheating (BEC) spinodal curve of  $h=\frac{3}{2}$,
\be
\ba{ll}
\sigma\simeq-2\varepsilon+{\frac{32}{\sqrt{\pi}\zeta(\frac{3}{2})}\varepsilon^{\frac{3}{2}}\,,}\quad&\quad
\varepsilon\simeq{\left[\frac{1}{32}\sqrt{\pi}\zeta(\frac{3}{2})\right]^{\frac{2}{3}}N^{-\frac{2}{3}}\,.}
\ea
\label{superheatse}
\ee
\end{list}
~\\

\begin{figure}[H]
\centering\includegraphics[width=100mm]{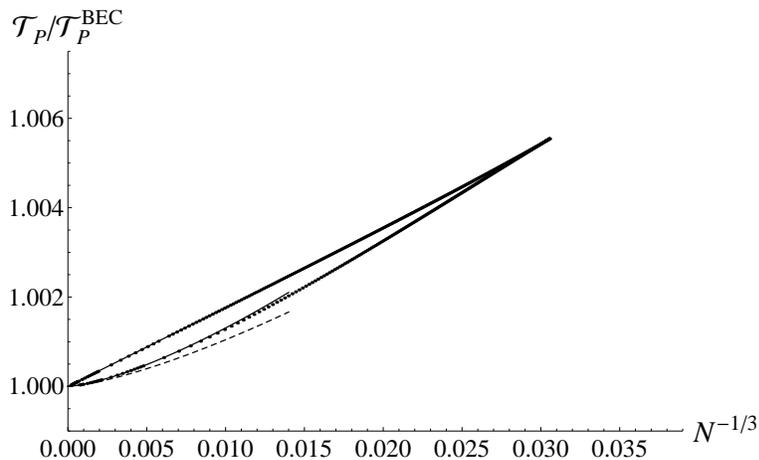}
\caption{The supercooling (lower)   and the superheating  (upper)  spinodal curves  on the $(N^{-1/3},{\Tp/\Tp^{\BEC}})$-plane.
The dotted curves   are from the numerical computations based on the exact formulas, (\ref {Nexpression2}), (\ref{defTp}), (\ref{Phispinodal}). The solid lines correspond to our  analytic approximate solutions,  (\ref{MAINsupercooling}), (\ref{MAINsuperheating}) for large $N$. The dashed  line denotes the  less-accurate  analytic approximation for the supercooling  curve without the  double logarithmic correction, \textit{i.e.~}`$\ln\ln N$', in  (\ref{MAINsupercooling}), \textit{c.f.~}\cite{ChabaPathria}. A pair of spinodal curves  emerge at $N_{\rm{c}}\simeq 35130.3$ ($N^{-1/3}_{\rm{c}}\simeq 0.0305332$) and start to converge  from  $N_{\mathrm{{\scriptscriptstyle{MAX}}}}\simeq 1.43056\times 10^{6}$  ($N_{\mathrm{{\scriptscriptstyle{MAX}}}}^{-1/3}\simeq 0.00887491$) toward the large $N$  limit.    }
\label{FIGspinodal}
\end{figure}

\begin{figure}[H]
\centering\includegraphics[width=100mm]{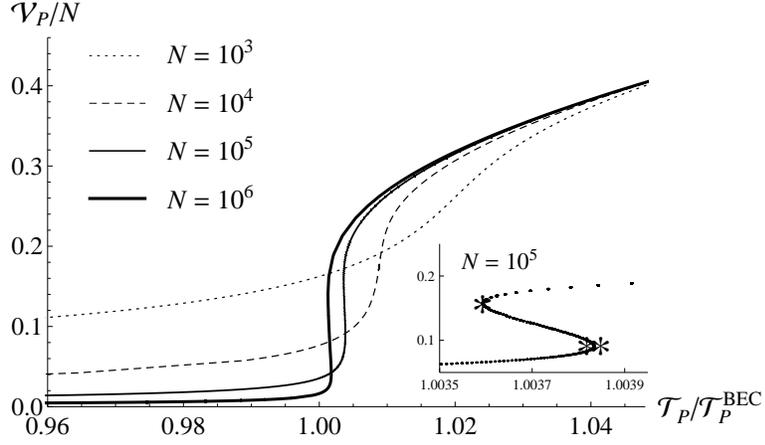}
\caption{$2D$ BEC in the position space:  if  $35131\leq N<\infty$, the  isobar zigzags on the $(\Tp/\Tp^{\BEC},{\Vp/N})$-plane with  the two turning points, supercooling\,$\ast$ (\ref{MAINsupercooling}) and superheating\,$\shp$ (\ref{MAINsuperheating}).  }
\label{FIGisobar}
\end{figure}

\begin{figure}[H]
\centering\includegraphics[width=100mm]{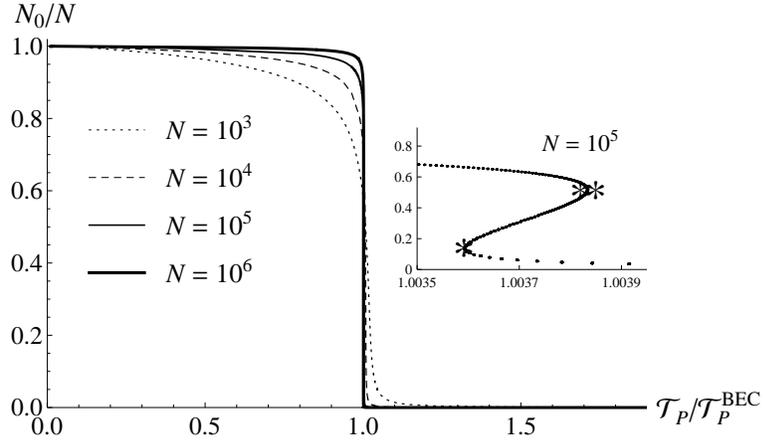}
\caption{$2D$ BEC in the momentum space: the number of the particles on the ground state zigzags   when  $35131\leq N<\infty$.}
\label{FIGd2bec}
\end{figure}

The above analytic approximation   agrees with the results obtained   by Chaba and Pathria~\cite{ChabaPathria}, where the double logarithmic terms were yet  neglected.  As shown  in Figures~\ref{FIGspinodalraw},\,\ref{FIGspinodal},  the   terms appear  crucial to match with the numerical computations. Besides   the double logarithmic  corrections, another novel contribution of  this work is to identify the two spinodal curves as the supercooling and the superheating turning points of the zigzagging isobar on $(T,V)$-plane\footnote{While superheated BEC has been  realized in  a recent  experiment~\cite{superheatedBEC},  our referring  of the other spinodal solution  as the `supercooling' may be arguable: simply calling it as `boiling point' is an alternative option~\cite{Hasok}.  Yet, in the present paper we stick to the convention of the precedents~\cite{7616PRA,7616relativistic,Jeon:2011nk}. },   see Figures~\ref{FIGisobar},\,\ref{FIGd2bec}.\\

In terms of the physical variables,  $N$, $\Tp$~(\ref{defTp}),  $\Vp$~(\ref{defVp}), $\Trho$~(\ref{defTrho}),  the supercooling and the superheating (BEC) points are as follows, \textit{c.f.~}\cite{ChabaPathria}.
\begin{list}{}
\item $\ast$ {\it{Supercooling  point\,}}:
\be
\ba{l}
\Tp^{\ast}/{\Tp^{\BEC}}\simeq 1+{\frac{3\sqrt{2}}{4\pi^{2}}\zeta(\frac{3}{2})}\left(\frac{\ln N+3\ln\ln N}{N}\right)^{\frac{1}{2}}\,,\\
\Vp^{\ast}\simeq {\sqrt{\frac{8\pi}{3}}}\frac{N}{\,\ln N+3\ln\ln N\,}\,,\\
\Trho^{\ast}\simeq\frac{8}{\pi}\left(\ln N+3\ln\ln N\right)^{-1}\,,\\
N_{0}/N\simeq{\frac{\sqrt{2}}{\zeta(\frac{3}{2})}}N^{-\frac{1}{2}}(\ln N+3\ln\ln N)^{\frac{3}{2}}\,.
\ea
\label{MAINsupercooling}
\ee

\item $\shp$ {\it{Superheating (BEC) point\,}}:   
\be
\ba{l}
\Tp^{\shp}/{\Tp^{\BEC}}\simeq 1
+{\frac{9}{16}\left[\frac{2\zeta^{4}(\frac{3}{2})}{\pi^{7}}\right]^{\frac{1}{3}}}
N^{-\frac{1}{3}}\,,\\
\Vp^{\shp}\simeq
{\frac{\,\pi^{\frac{7}{6}}}{\sqrt{24}}\left[\frac{32}{\zeta(\frac{3}{2})}\right]^{\frac{2}{3}}}N^{\frac{2}{3}}\,,\\
\Trho^{\shp}\simeq\left[\frac{1}{32}\sqrt{\pi}\zeta(\frac{3}{2})\right]^{-\frac{2}{3}}N^{-\frac{1}{3}}\,,\\
N_{0}/N\simeq 1-{\left[\frac{128\pi^{2}}{27\zeta^{2}(\frac{3}{2})}\right]^{\frac{1}{3}}}N^{-\frac{1}{3}}\ln N\,.
\ea
\label{MAINsuperheating}
\ee
\end{list}
\noindent Here ${\Tp^{\BEC}={\sqrt{{24}/{\pi^{3}}}}}$  denotes a dimensionless  constant which gives  the  $2D$  BEC critical temperature  in the large $N$  limit~\cite{Imry}\,:
\be
{\kB T^{\BEC}=\hbar\sqrt{\frac{12}{\pi m}P\,} \,.}
\label{TBECc}
\ee
As we show shortly, this formula  holds universally even  for the  $2D$   harmonic potential.

The above results lead to the \textit{phase diagram} of the $2D$ ideal Bose gas on the $(T,P)$-plane: under   arbitrarily fixed   pressure, $P$, the state is \textit{condensate} if  
\be
{
T<T^{\ast}=\frac{\hbar}{\kB}\sqrt{\frac{12}{\pi m}P\,}\left[1+{\frac{3\sqrt{2}}{4\pi^{2}}\zeta(\textstyle{\frac{3}{2}})}\sqrt{\frac{\ln N+3\ln\ln N}{N}}\,\right]\,,}
\ee 
and  becomes  \textit{gas}  if
\be
{
T\geq T^{\shp}=\frac{\hbar}{\kB}\sqrt{\frac{12}{\pi m}P\,}\left[ 1
+{\frac{9}{16}\sqrt[3]{\frac{2\zeta^{4}(\frac{3}{2})}{\pi^{7}N}}}\,\right]\,.}
\ee
Note that $T^{\shp}>T^{\ast}$. Between the two temperatures, three states may coexist, including  one unstable.  In the large $N$ limit, the gap, $T^{\shp}-T^{\ast}=\cO(N^{-1/3})$,  becomes negligible  and  the two spinodal curves  converge to a single line on the $(T,P)$-plane. Further the density of the superheated BEC is as high as  ${N/\Vp^{\shp}}= \cO(N^{1/3})$ and   diverges in the large $N$   limit.

In contrast to  the $3D$ ideal Bose gas~\cite{7616PRA,Jeon:2011nk},  the volume per particle at the supercooling point is not constant but depends inversely on $\ln N$  \textit{i.e.~}$\Vp^{\ast}/N=\cO(\frac{1}{\ln N})$. Hence  both $\Vp^{\ast}/N$ and $\Vp^{\shp}/N$ vanish in the large $N$  limit and therefore, once drawn on the $(\Tp,\Vp/N)$-plane  the superheating and the supercooling points  converge to the single point, $(\Tp^{\BEC},0)$. In order to see their separation even in the large $N$ limit, it is necessary  to consider a rather unusual  variable,  $N/(V\ln N)$, instead of  the density, $N/V$. Otherwise the conventional thermodynamic limit with fixed density fails to capture  the fine structure of the large $N$ behaviors of the ideal gas. \\


\section{Consistency with a $2D$ harmonic trap}
 For an arbitrary gas  subject to a  $2D$   harmonic potential, $V(r)=\half m\omega^{2}r^{2}$,  let us define  $N(r)$  to be the number of particles within the radius $r$  and  $P(r)$ to be    the radially varying   pressure. They assume    boundary values,  ${N(0)=0}$, ${N(\infty)=N}$,  $P(\infty)=0$, and satisfy   the Newtonian equilibrium condition,  balancing the  pressure gradient and the  harmonic force,
\be
2\pi r P^{\prime}(r)=-mN^{\prime}(r) \omega^{2}r\,.
\ee
It follows then
\be
{P(r)=\frac{m\omega^{2}}{2\pi}\left[N-N(r)\right]\,. }
\ee
In particular, at the center, $r=0$, where BEC is typically observed in experiment,   we have  
\be
P(0)=\frac{m\omega^{2}}{2\pi}N\,.
\label{centralP}
\ee 
Substituting this   into Eq.(\ref{TBECc}),  we recover  precisely -- and satisfactorily --      the known BEC critical  temperature for the  $2D$  harmonic trap~\cite{HadzibabicDalibard,Mullin97,Colloquium},
\be
{\kB T^{\BEC}=\frac{\hbar\omega}{\pi}\sqrt{6N\,}\,.}
\ee
It is worth while to note that the value of the central pressure~(\ref{centralP}) is  true not only for the ideal gas but also for real gases,  regardless of  temperature,  species  and  interactions.  The derivation above is independent of them. Thus,  at the center of the harmonic potential the pressure is kept fixed and  the Bose-Einstein condensation can take place discretely. 
~\\

\section{Discussion}
To summarize,  when $N\geq N_{c}=35131$, the ideal Bose gas  confined in a   $2D$   box   reveals   thermodynamic instability or   a pair of spinodal curves, supercooling and  superheating (BEC). Evading  the MWH no-go theorem, the gas condenses  discretely at   finite temperature under constant pressure    in both the momentum  and  the position   spaces.  As in $3D$~\cite{7616PRA,7616relativistic,Jeon:2011nk}, this is   an emergent  phenomenon which  finitely many  bosonic identical particles can feature, without assuming the large $N$ limit.

  The ideal gas  represents  the leading order behavior of any (weakly) interacting real gases.  It is natural to expect that small  interactions should  deform the shape of the spinodal curve depicted in Figure~\ref{FIGspinodal} but  hard  to imagine that such interactions will make the spinodal curve completely  disappear. Thus,  our result  should   provide  a novel    theoretical foundation for  a two-dimensional  discrete  phase transition of real gases.

Some reasons how our result evades the MWH  theorem are as follows.  Firstly, the theorem assumes the density to be  finite. Yet,  the density of the superheated BEC we have obtained  diverges in the large $N$ limit. Therefore,    the  theorem is inapplicable  to our case, see also \cite{Mullin97} for related discussion.   Secondly, the theorem concerns  the infrared divergence of the momentum integral, $\int{\rm d}^{s}k\,\frac{1}{\,k^{2}\,}$,  which diverges in one and  two  spatial dimensions, \textit{i.e.~}$s=1,2$, (see the discussion around Eqs.(18,19) in \cite{HohPR}). However,  at the quantum level, the discrete quantum states should convert the  integral to a   sum over  the countable   states. The sum is finite  and hence the theorem can be circumvented.  Repulsive interactions may well prevent density from diverging as expected from real gases. As discussed above, only if the interaction does not alter the existence of the spinodal  curve, a discrete phase transition should persist.  Then,  the second   reason should survive to  explain why the MWH theorem may  not hold for a real quantum system of  discrete  spectrum.

Although a two-dimensional  square box potential has been realized in a recent experiment~\cite{14064073}, it should be hard to impose the constant pressure  condition especially on small box systems.  On the other hand,  we have shown that the  pressure at the center of the harmonic potential remains  fixed for both the ideal and  real gases, (\ref{centralP}). Our result  then seems to indicate   that   experimentally observed Bose-Einstein  condensations of harmonically trapped  atomic gases  are a  first-order  phase transition which involves  a discrete change of the density  at least   at the center of the trap, provided there are sufficiently many particles.\footnote{We thank one of the anonymous referees for helping us to make this assertion in the present  revised version of the  manuscript.}

In this work we have focused on the grand canonical ensemble.  Analyzing instead  the canonical ensemble  will require more computational power and  most likely suggest  a different  value of the critical number: for example in $3D$,   $N_{c}=7616$ (canonical)~\cite{7616PRA} or $N_{c}=14393$ (grand canonical)~\cite{Jeon:2011nk}. Yet, in our opinion, what matters is  the existence of such a definite   number,  alternative to infinity,    which  may   answer  to the   question,  \textit{How many is different?}~\cite{Park:2013gpa}. For larger values of $N$  the differences due to   different  choices of the ensembles  are  expected to be  anyhow negligible. In $3D$,  the canonical ensemble results agree with  those of the grand canonical ensemble within $0.1\%$ error when  $N=10^{5}$ or $10^{6}$, \textit{c.f.~}the TABLE~I in   Ref.\cite{Jeon:2011nk}.

As computable from our analytic solutions,   (\ref{MAINsupercooling}), (\ref{MAINsuperheating}),  the gap between the supercooling and the superheating temperatures, $\Tp^{\shp}-\Tp^{\ast}$, becomes  maximal when the number of particles  is equal to  $N_{\mathrm{{\scriptscriptstyle{MAX}}}}\simeq 1.43056\times 10^{6}$. 
This also  agrees with the numerical result  shown  in  Figure~\ref{FIGspinodal}.   In a way,  the  two numbers, $N_{c}$ and $N_{\mathrm{{\scriptscriptstyle{MAX}}}}$, enable   us to divide  the Bose gas system  into three quantum realms: 
\begin{enumerate}
\item   \textit{microscopic\,} for\, ${1\leq N<N_{c}}$, 
\item  \textit{mesoscopic\,} for\, ${N_{c}\leq N\leq N_{\mathrm{{\scriptscriptstyle{MAX}}}}}$, 
\item \textit{macroscopic\,}  for\, ${N_{\mathrm{{\scriptscriptstyle{MAX}}}}<N\leq\infty}$.  
\end{enumerate}

Lastly,  when  $N=N_{c}$  the volume expansion  ratio,
\be
{
\Vp^{\ast}/\Vp^{\shp}\simeq \left[\frac{\zeta(\frac{3}{2})}{\sqrt{2}\pi}\right]^{\frac{2}{3}}\!
\frac{N^{\frac{1}{3}}}{\ln N+3\ln\ln N}\simeq 0.701855\times\!\frac{N^{\frac{1}{3}}}{\ln N+3\ln\ln N}\,,}
\label{VolumeExp}
\ee
becomes $1.31267$ which is `for consistency' of order unity.\\

\ack
We would like to thank  Chuan-Tsung Chan, Hasok Chang  and Yong-il Shin  for helpful  comments. This work was  supported by  the National Research Foundation of Korea (NRF)   with the Grants,  2012R1A2A2A02046739 and 2013R1A1A1A05005747.\\

\newpage

\appendix
\section{Analytic Approximation\label{AppendixA}}
To solve the spinodal condition~(\ref{Phispinodal}) and to compute the number of particles~(\ref{Nexpression2}),  we  henceforth focus  on
\be
\ba{l}
\partial_{\sigma}\ln\cZ=\sum_{\vn\in\mN^{2}}\,\frac{1}{1-e^{\varepsilon\vn^{2}+\sigma}}\,,\\
\varepsilon^{-1}\partial_{\varepsilon}\ln\cZ=\sum_{\vn\in\mN^{2}}\,\frac{\varepsilon^{-1}\vn^{2}}{1-e^{\varepsilon\vn^{2}+\sigma}}\,,\\
\partial^{2}_{\sigma}\ln\cZ=\sum_{\vn\in\mN^{2}}\,\frac{1}{4}\sinh^{-2}\left(\frac{\varepsilon\vn^{2}+\sigma}{2}\right)\,,\\
\partial_{\varepsilon}\partial_{\sigma}\ln\cZ=\sum_{\vn\in\mN^{2}}\,\frac{1}{4}\vn^{2}\sinh^{-2}\left(\frac{\varepsilon\vn^{2}+\sigma}{2}\right)\,,\\
\partial^{2}_{\varepsilon}\ln\cZ=\sum_{\vn\in\mN^{2}}\,\frac{1}{4}(\vn^{2})^{2}\sinh^{-2}\left(\frac{\varepsilon\vn^{2}+\sigma}{2}\right)\,.
\ea
\label{ppZ}
\ee
Our  computational scheme to obtain the  analytic solution of (\ref{Phispinodal}) is, based on Ref.\cite{Jeon:2011nk},    as follows.

\begin{enumerate}
\item  {Separate the lattice sum into two parts introducing  an arbitrary  cutoff,} $\Lambda=5,8,10,13,17,\cdots$  $(\Lambda>2)$,
\be
{\sum_{\vn\in\mN^{2}}f(\varepsilon\vn^{2})
=\sum_{\vn^{2}<\Lambda}f(\varepsilon\vn^{2})
+\sum_{\vn^{2}\geq\Lambda}f(\varepsilon\vn^{2})\,.}
\label{scheme1}
\ee
\item {Approximate the last sum  by an integral}, 
\be
{\quad{{\sum_{\vn^{2}\geq\Lambda}}}
f(\varepsilon\vn^{2})\simeq
{{\int_{\Lambda\varepsilon}^{\infty}}{\,\rd x\,}}
\left[{\frac{\pi}{4}\varepsilon^{-1}-{\frac{1}{2}}(\varepsilon x)^{-\frac{1}{2}}}\right]f(x)\,.}
\label{schemeMain}
\ee
This formula  follows from an identity,
\[
{
\sum_{\vn\in\mN^{2},\vn^{2}\geq\Lambda}=\frac{1}{4}\sum_{\vn\in\mZ^{2},\vn^{2}\geq\Lambda}-\frac{1}{2}
\sum_{n\in\mZ,n^{2}\geq\Lambda}\,,}
\]
and the   integral approximation  of the right hand side~\cite{Grossmann95}.
\item {Assume an ansatz  with two positive quantities:}
\be
\ba{lll}
\sigma\simeq -2\varepsilon +g\varepsilon^{h}\,,~&~g>0\,,~&~h>0\,.
\ea
\label{ansatz}
\ee
{While we put} $h$ {to be a constant, we allow} $g$ {to  depend possibly on} `$\ln\varepsilon$'. The appearance of  logarithmic dependency  is a novel feature in   $2D$  compared with   $3D$~\cite{Jeon:2011nk}.  The number of the particles on the ground state  is now,
\be
N_{0}\simeq g^{-1}\varepsilon^{-h}\,.
\ee
{We further set} $h\neq 1$, {as it agrees  with  the numerical results and simplifies our algebraic analysis.}

\item Expand each quantity  in (\ref{ppZ})  in powers of $\varepsilon$. For consistency, we should   trust only the singular terms  which are    insensitive  to the cutoff, $\Lambda$. We shall see that   for each quantity,    at least  first two   leading  singular powers  are   $\Lambda$-independent.

\end{enumerate}
Our scheme  implies then, as $\varepsilon\rightarrow 0^{+}$,
\be
\sigma\simeq\left\{\ba{clc}
g\varepsilon^{h}>0&\mbox{for}&h<1\\
-2\varepsilon<0&\mbox{for}&h>1\,,\ea\right.
\ee
and
\be
\varepsilon\vn^{2}+\sigma\simeq
\left\{\ba{clccl}
g\varepsilon^{h}&\mbox{for}&h<1&\mbox{or}&\vn^{2}=2\\
(\vn^{2}-2)\varepsilon&\mbox{for}&h>1&\mbox{and}&\vn^{2}>2\,.\ea\right.
\label{sh}
\ee
We  set some constants:
\be
\ba{l}
a_{s}:=\dis{\int_{0}^{\infty}{\rd x}
{\frac{x^{s}}{e^{x}-1}}=\Gamma(s+1)\zeta(s+1)\,,}\\
b_{t}:=\dis{\int_{0}^{\infty}{\rd x}
\frac{x^{t}}{e^{x}+1}= (1-\frac{1}{2^{t}})\Gamma(t+1)\zeta(t+1)\,,}
\ea
\label{asbt}
\ee
and consider the   associated  $\varepsilon$-dependent integrals:
\be
\ba{ll}
\alpha_{s}:=\dis{
\int_{\Lambda\varepsilon}^{\infty}\rd x\,\frac{x^{s}}{e^{x+\sigma}-1}\,,}~~&~~\dis{
\beta_{t}:={\int_{\Lambda\varepsilon}^{\infty}\rd x\,
\frac{x^{t}}{e^{x+\sigma}+1}}\,.}
\ea
\label{abo}
\ee
For $s>0$, $t>-1$, the constants, $a_{s},b_{t}$, are finite. Hence   neglecting nonsingular terms we may estimate
\be
\ba{ll}
\alpha_{s}\simeq a_{s}~~\mbox{for}~~s>0\,,~~~&~~~
\beta_{t}\simeq b_{t}~~\mbox{for}~~t>-1\,.
\ea
\ee
It is then straightforward to see
\be
\varepsilon^{-1}\partial_{\varepsilon}\ln\cZ\simeq
-2g^{-1}\langle\varepsilon^{-(h+1)}\rangle_{-\frac{5}{2}}-\frac{\pi}{4}a_{1}\varepsilon^{-3}+\frac{1}{2}a_{\frac{1}{2}}\varepsilon^{-\frac{5}{2}}\,,
\label{fpe}
\ee
where, as introduced in \cite{Jeon:2011nk}, $\langle\varepsilon^{-(h+1)}\rangle_{-\frac{5}{2}}$ is equal to $\varepsilon^{-(h+1)}$ if $h+1\geq\frac{5}{2}$, otherwise  it is zero. \\

We also have from (\ref{sh}), (\ref{a0}),
\be
\ba{ll}
\alpha_{0}&=-\ln(1-e^{-\Lambda\varepsilon-\sigma})\\
{}&
\simeq-\ln\left((\Lambda{-2})\varepsilon+g\varepsilon^{h}\right)
\simeq\left\{\ba{lll}
-h\ln\varepsilon~&\mbox{for}&h<1\\
-\ln\varepsilon~&\mbox{for}&h>1\,,
\ea
\right.
\ea
\label{alpha0}
\ee
and from (\ref{1e1}),   (\ref{am0p5}), (\ref{infsumb}),
\be
\alpha_{-\frac{1}{2}}
\simeq\left\{\ba{lll}
\pi g^{-\frac{1}{2}}\varepsilon^{-\frac{h}{2}}-(\sqrt{2}+1)b_{-\frac{1}{2}}&~\mbox{for}&h<1\\
\frac{1}{\sqrt{2}}\varepsilon^{-\frac{1}{2}}
\ln\left|\frac{\sqrt{\Lambda}+\sqrt{2}}{\sqrt{\Lambda}-\sqrt{2}}\right|
-(\sqrt{2}+1)b_{-\frac{1}{2}}&~\mbox{for}&h>1\,.
\ea
\right.
\label{alpham0p5}
\ee
From  these,  the  first  two reliable leading  singular  terms  for the number of particles, 
$N$, given in (\ref{Nexpression2}),   are
\be
-\partial_{\sigma}\ln\cZ\simeq\left\{\ba{lll}
-\frac{\pi}{4}\varepsilon^{-1}\ln(g\varepsilon^{h})-\frac{\pi}{2}g^{-\frac{1}{2}}\varepsilon^{-\frac{1+h}{2}}&\!\mbox{for}\!&\!h<1\\
g^{-1}\varepsilon^{-h}-\frac{\pi}{4}\varepsilon^{-1}\ln\varepsilon&\!\mbox{for}\!&\!h>1\,.\ea\right.
\label{fps}
\ee
In particular, we note that $N_{0}\simeq g^{-1}\varepsilon^{-h}$ is significant in $N$ (indicating  BEC)   only for $h>1$.

Now, we turn to the computations of the second order derivatives in (\ref{ppZ}), which in part requires us to consider
\be
\omega_{r}:=\int_{\Lambda\varepsilon}^{\infty}\rd x\, 
\textstyle{\frac{1}{4}x^{r}\sinh^{-2}(\frac{x+\sigma}{2})\,.}
\ee
For $r>1$, an integration by parts   with trivial boundary contribution  gives a simple relation between $\omega_{r}$ and $\alpha_{s}$,
\be
\omega_{r}=\int_{\Lambda\varepsilon}^{\infty}\rd x\,x^{r}\frac{{\rm d}~}{{\rm d}x}\!\left(1-e^{x+\sigma}\right)^{-1}\simeq r\alpha_{r-1}\,,
\ee
such that
\be
\ba{ll}
\omega_{2}\simeq 2\alpha_{1}\,,\quad\quad&\quad\quad
\omega_{\frac{3}{2}}\simeq \frac{3}{2}\alpha_{\frac{1}{2}}\,.
\ea
\ee
It follows then
\be
\ba{l}
\partial_{\varepsilon}^{2}\ln\cZ\simeq 4g^{-2}\langle\varepsilon^{-2h}\rangle_{-\frac{5}{2}}+
\frac{\pi}{2}a_{1}\varepsilon^{-3}-\frac{3}{4}a_{\frac{1}{2}}\varepsilon^{-\frac{5}{2}}\,.
\ea
\label{fpepe}
\ee
For $r=1$ we perform the same partial integration,  this time  receiving  nontrivial yet non-singular  boundary contribution.   We  obtain with (\ref{a0}),
\be
{\omega_{1}=-\ln(1-e^{-\Lambda\varepsilon-\sigma})+\frac{\Lambda\varepsilon}{e^{\Lambda\varepsilon+\sigma}-1}\simeq\alpha_{0}\,.}
\ee
For $r=\frac{1}{2}$, using (\ref{1e2}),   (\ref{o0p5}),  (\ref{infsumb}),  we  get
\be
\omega_{\frac{1}{2}}\simeq\left\{
\ba{lll}
\half\alpha_{-\frac{1}{2}}~&\mbox{for}&h<1\\
\frac{\sqrt{\Lambda}}{\Lambda-2}\varepsilon^{-\frac{1}{2}}+\half\alpha_{-\frac{1}{2}}~&\mbox{for}&h>1\,.\ea
\right.
\ee
Our scheme then gives 
\be
\partial_{\varepsilon}\partial_{\sigma}\ln\cZ\simeq
\left\{
\ba{lll}
-\frac{\pi}{4}\varepsilon^{-2}\ln(g\varepsilon^{h})-\frac{\pi}{4}g^{-\frac{1}{2}}\varepsilon^{-\frac{3+h}{2}}~&\mbox{for}&h<1\\
2g^{-2}\varepsilon^{-2h}-\frac{\pi}{4}\varepsilon^{-2}\ln\varepsilon~&\mbox{for}&h>1\,.\ea
\right.
\label{fpeps}
\ee
Finally, from (\ref{1e2}), (\ref{om0p5}), we have
\be
\ba{ll}
\partial_{\sigma}^{2}\ln\cZ&=\sum_{\vn^{2}\in\mN^{2}}\left[\frac{1}{(\varepsilon\vn^{2}+\sigma)^{2}}
-\sum_{k=2}^{\infty}\frac{1}{4^{k}}\cosh^{-2}(\frac{\varepsilon\vn^{2}+\sigma}{2^{k}})\right]\\
{}&\simeq\sum_{\vn^{2}<\Lambda}\frac{1}{(\varepsilon\vn^{2}+\sigma)^{2}}
+{\dis{\int}}_{\Lambda\varepsilon}^{\infty}{\rd x\,}
\frac{\frac{\pi}{4}\varepsilon^{-1}-{\frac{1}{2}}(\varepsilon x)^{-\frac{1}{2}}}{(x+\sigma)^{2}}-\frac{\pi}{8}\varepsilon^{-1}\\
{}&\simeq\left\{
\ba{lll}
\frac{\pi}{4}g^{-1}\varepsilon^{-1-h}
-\frac{\pi}{8}\varepsilon^{-1}
&\mbox{for}&0<h<\frac{1}{3}\\
\frac{\pi}{4}g^{-1}\varepsilon^{-\frac{4}{3}}
-\frac{\pi}{4}(g^{-\frac{3}{2}}+\frac{1}{2})\varepsilon^{-1}
&\mbox{for}&h=\frac{1}{3}\\
\frac{\pi}{4}g^{-1}\varepsilon^{-1-h}
-\frac{\pi}{4}g^{-\frac{3}{2}}\varepsilon^{-\frac{1+3h}{2}}
&\mbox{for}&\frac{1}{3}<h<1\\
g^{-2}\varepsilon^{-2h}+\sum_{\vn^{2}>2}\frac{1}{(\vn^{2}-2)^{2}}\varepsilon^{-2}&\mbox{for}&h>1\,.
\ea
\right.
\ea
\label{fpsps}
\ee
Here for  $h>1$,  we have chosen $\Lambda\rightarrow\infty$.

The numerical values of the constants are
\be
\ba{ll}
a_{\frac{1}{2}}=\frac{\sqrt{\pi}}{2}\zeta(\frac{3}{2})\simeq 2.31516\,,\quad&\quad
a_{1}=\frac{\,\pi^{2}}{6}\simeq 1.64493\,,\\
\multicolumn{2}{c}{
\sum_{\vn^{2}>2}\frac{1}{(\vn^{2}-2)^{2}}\simeq 0.351699 \,.}
\ea
\ee

Having the key expressions,  (\ref{fpe}), (\ref{fps}), (\ref{fpepe}), (\ref{fpeps}), (\ref{fpsps}),    we are now ready to   solve the spinodal  condition of Eq.(\ref{Phispinodal}).   We consider      eight possible  cases separately:  
\[
\ba{cccc}
{0<h<\frac{1}{3}}\,,&\quad{h=\frac{1}{3}}\,,&\quad{\frac{1}{3}< h<1}\,,&\quad{1<h<\frac{5}{4}}\,,\\
{h=\frac{5}{4}}\,,&\quad{\frac{5}{4}<h<\frac{3}{2}}\,,&\quad{h=\frac{3}{2}}\,,&\quad{\frac{3}{2}<h}\,.
\ea
\] 
As our  ansatz~(\ref{ansatz}) contains  a single unknown  term,  we demand at least the leading power in $\Phi$ should be canceled out in a nontrivial manner.      It is straightforward to check  that only  the two values, $h=\frac{1}{2}$ and $h=\frac{3}{2}$, admit   solutions, leading to  (\ref{supercoolse}) and (\ref{superheatse}) respectively. As shown in  Figure~\ref{FIGspinodalraw}, they are in good agreement with the numerical
 result.

  \newpage

\section{Useful identities and integrals, \textit{c.f.}~\cite{Jeon:2011nk}}
\noindent
\be
{(1-e^{x})^{-1}=-x^{-1}+{\sum_{n=1}^{\infty}}\,\frac{1}{\,2^{n}}
(1+e^{\frac{x}{2^{n}}})^{-1}}\,.
\label{1e1}
\ee
\be
\ba{ll}
{\frac{1}{4}}\sinh^{-2}\left(\frac{x}{2}\right)
{}\!\!\!\!&=x^{-2}\,-{\sum_{k=2}^{\infty}}\,\frac{1}{\,4^{k}}\cosh^{-2}\left(\frac{x}{2^{k}}\right)\\
&=x^{-2}\,+{\sum_{n=1}^{\infty}}\frac{1}{\,2^{n}}
\frac{{\rm d}~}{{\rm d}x}\!\left(1+e^{\frac{x}{2^{n}}}\right)^{-1}\,.
\ea
\label{1e2}
\ee
\be
{
\int\rd x \,\left(e^{x+\sigma}-1\right)^{-1}=\ln(1-e^{-x-\sigma})}\,.
\label{a0}
\ee
\be
{
\int\rd x\,\left[\sqrt{x}(x+\sigma)\right]^{-1}}
=\left\{\ba{lll}
\frac{2}{\sqrt{\sigma}}\arctan(\sqrt{x/\sigma})&\mbox{~for}&\sigma>0\\
\frac{1}{\sqrt{-\sigma}}\ln\left|\frac{\sqrt{-x/\sigma}-1}{
\sqrt{-x/\sigma}+1}\right|&\mbox{~for}&\sigma<0\,.
\ea\right.
\label{am0p5}
\ee
\be
\ba{l}
\dis\int\rd x\left[\sqrt{x}(x+\sigma)^{2}\right]^{-1}\\
=\left\{\ba{ll}
\sigma^{-1}\left[\frac{\sqrt{x}}{x+\sigma}+\frac{1}{\sqrt{\sigma}}\arctan(\sqrt{x/\sigma})\right]&\mbox{~for~}
\sigma>0\\
\sigma^{-1}\left[\frac{\sqrt{x}}{x+\sigma}+\frac{1}{2\sqrt{-\sigma}}\ln\left|\frac{\sqrt{-x/\sigma}-1}{
\sqrt{-x/\sigma}+1}\right|\right]&\mbox{~for~}\sigma<0\,.
\ea\right.
\ea
\label{om0p5}
\ee
\be
\ba{l}
\dis\int\rd x\,\frac{\sqrt{x}}{(x+\sigma)^{2}}\\
=\left\{\!\ba{ll}
-\frac{\sqrt{x}}{x+\sigma}+\frac{1}{\sqrt{\sigma}}\arctan(\sqrt{x/\sigma})&\mbox{~for~}\sigma>0\\
-\frac{\sqrt{x}}{x+\sigma}+\frac{1}{2\sqrt{-\sigma}}\ln\left|\frac{\sqrt{-x/\sigma}-1}{
\sqrt{-x/\sigma}+1}\right|&\mbox{~for~}\sigma<0\,.
\ea\right.
\ea
\label{o0p5}
\ee
For $-1<t<0$, the following sum converges,
\be
{\sum_{n=1}^{\infty}\frac{1}{2^{n}}\int_{0}^{\infty}{\rd x}\,
{x^{t}}\left(e^{\frac{x}{2^{n}}}+1\right)^{-1}=\frac{b_{t}}{2^{-t}-1}}\,.
\label{infsumb}
\ee


\newpage

\section*{References}
\begin{thebibliography}{99}
\bibitem{MerPRL}
N. D. Mermin and H. Wagner,  
``Absence of Ferromagnetism or Antiferromagnetism in One- or Two-Dimensional Isotropic Heisenberg Models,"
\textit{Phys. Rev. Lett.} {\bf 17} 1133 (1966).

\bibitem{HohPR}
P. C. Hohenberg, 
``Existence of Long-Range Order in One and Two Dimensions,"
\textit{Phys. Rev.} {\bf 158} 383 (1967).


\bibitem{BerSPJ}
V. L. Berezinskii,
``Destruction of Long-range Order in One-dimensional and Two-dimensional Systems Possessing a Continuous Symmetry Group. II. Quantum Systems,"
\textit{Sov. Phys. JETP.} {\bf 34} 610 (1972).

\bibitem{KosJPC}
J. M. Kosterlitz and D. J. Thouless,
``Ordering, metastability and phase transitions in two-dimensional systems,"
\textit{J. Phys. C} {\bf 6} 1181 (1973).




\bibitem{HadzibabicDalibard}
Z. Hadzibabic and J. Dalibard,   
``Two-dimensional Bose fluids: An atomic physics perspective,''  
\textit{Rivista del Nuovo Cimento} {\bf 34} 389 (2011).


\bibitem{BagnatoKleppner}
V. Bagnato and D. Kleppner,  
``Bose-Einstein condensation in low-dimensional traps,''
  \textit{Phys. Rev. A} {\bf 44}  7439-7441 (1991). 

\bibitem{Mullin97}
W. J. Mullin,  
``Bose-Einstein condensation in a harmonic potential,''
\textit{J. Low Temp. Phys.} {\bf 106} 615 (1997).




\bibitem{Colloquium}
A. Posazhennikova,  
``Colloquium:~Weakly interacting, dilute Bose gases in  2D,'' 
\textit{Rev. Mod. Phys.} {\bf 78} 1111 (2006).


\bibitem{Anderson} 
P. W. Anderson,
``More Is Different,"   \textit{Science} 177, 393 (1972). 

\bibitem{Park:2013gpa}
  J.-H.~Park,
  ``How many is different? Answer from ideal Bose gas,''
  \textit{J. Phys. Conf. Ser.}  {\bf 490} 012018 (2014), 
  [arXiv:1310.5580 [cond-mat.stat-mech]].

\bibitem{YangLee}  
C. N. Yang and T. D. Lee, 
``Statistical Theory of Equations of State and Phase Transitions. I. Theory of Condensation," \textit{Phys. Rev.} {\bf 87}  404 (1952); 
``Statistical Theory of Equations of State and Phase Transitions. II. Lattice Gas and Ising Model,"  \textit{Phys. Rev.} {\bf 87} 410 (1952).

\bibitem{Kadanoff} 
L. P. Kadanoff, 
``More is the Same; Phase Transitions and Mean Field Theories,"
\textit{J. Stat. Phys.} 137, 777 (2009).

\bibitem{London} 
F. London, \textit{Superfluids} Dover Publications, Inc.  (1954),   Vol. II,  Part 7.




\bibitem{spinodalnuclear}
P. Chomaz,  M. Colonna  and J. Randrup,  
``Nuclear spinodal fragmentation,"  
\textit{Phys. Rep.} {\bf  389} 263 (2004).

\bibitem{Kardar}
M. Kardar, 
\textit{Statistical Physics of Particles} 
Cambridge University Press, (2007).


\bibitem{Yukalov}
V. I. Yukalov,   
``Basics of Bose-Einstein Condensation,''   
\textit{Phys. Part. Nucl.} {\bf 42} 460 (2011).


\bibitem{Imry}
Y. Imry, D. J. Bergman and L. Gunther,  
 ``Bose-Einstein Condensation in Two-Dimensional Systems,"
\textit{Low Temperature Physics-LT 13} Springer US, (1974), Vol. 1, p. 80.




\bibitem{ChabaPathria}
A. N. Chaba and R. K. Pathria,  
 ``Bose-Einstein condensation in a two-dimensional system at constant pressure,"
  \textit{Phys. Rev. B} {\bf 12} 3697 (1975).



\bibitem{7616PRA}
J.-H. Park and  S.-W. Kim,  
``Thermodynamic instability and first-order phase transition in an ideal Bose gas,"
\textit{Phys. Rev. A} {\bf 81} 063636 (2010).

\bibitem{7616relativistic}
J.-H. Park and  S.-W. Kim,  
``Existence of a critical point in the phase diagram of ideal relativistic neutral  Bose gas,"
\textit{New J. Phys.} {\bf 13} 033003  (2011).

\bibitem{Jeon:2011nk}
I.~Jeon, S.-W.~Kim and J.-H.~Park,  
 ``Isobar of an ideal Bose gas within the grand canonical ensemble,''
  \textit{Phys. Rev. A} {\bf 84} 023636 (2011).







\bibitem{14064073}
L. Corman \textit{et al.},  
``Quench-induced supercurrents in an annular  Bose gas," 
\textit{Phys. Rev. Lett.} {\bf 113} 135320 (2014).




\bibitem{3Dharmonic}
S.  Grossmann and M. Holthaus,
``On Bose-Einstein condensation in harmonic traps," \textit{Phys. Lett. A} {\bf 208}  188 (1995).

\bibitem{Grossmann95}
S.  Grossmann and M. Holthaus,
``Bose-Einstein condensation in a cavity," \textit{Z. Phys. B} {\bf 97}  319 (1995).



\bibitem{superheatedBEC}  
A. L. Gaunt,	 R. J. Fletcher,	R. P. Smith and	Z. Hadzibabic,
``A superheated Bose-condensed gas,''
\textit{Nature Phys.} {\bf 9} 271-274 (2013).


\bibitem{Hasok} Hasok Chang, \textit{private communication.}
\end{thebibliography}
\end{document}